\documentstyle[aps]{revtex}
\begin{document}
 \draft
\def\be{\begin{equation}}
\def\ee{\end{equation}}
\def\ba{\begin{eqnarray}}
\def\ea{\end{eqnarray}}
\def\bq{\begin{quote}}
\def\eq{\end{quote}}
\def\PL{{ \it Phys. Lett.} }
\def\PRL{{\it Phys. Rev. Lett.} }
\def\NP{{\it Nucl. Phys.} }
\def\PR{{\it Phys. Rev.} }
\def\MPL{{\it Mod. Phys. Lett.} }
\def\IJMP{{\it Int. J. Mod .Phys.} }
\newcommand{\labell}[1]{\label{#1}\qquad_{#1}} %{\label{#1}} %
\newcommand{\labels}[1]{\vskip-2ex$_{#1}$\label{#1}} %{\label{#1}} %

\twocolumn[\hsize\textwidth\columnwidth\hsize\csname
@twocolumnfalse\endcsname
\preprint{CERN-TH-2000-038\\CWRU-P1-00\\SU-ITP-00/05\\
hep-ph/0002001\\ January 2000}
\date{\today}
\title{Compact Hyperbolic Extra Dimensions: Branes, Kaluza-Klein Modes and Cosmology}
\author{Nemanja Kaloper$^1$, John March-Russell$^2$, Glenn D. Starkman$^3$,
Mark Trodden$^3$}
\vskip.5cm
\address{{\em $^1$ Department of Physics, Stanford University, Stanford, CA
94305, USA}\\
{\em $^2$ Theory Division, CERN, CH-1211, Geneva 23,
Switzerland}\\
{\em $^3$
Department of Physics, Case Western Reserve University,
Cleveland, OH 44106-7079, USA}\\}
\maketitle
\begin{abstract}
We reconsider theories with low gravitational (or string) scale $M_*$ where Newton's
constant is generated via new large-volume spatial dimensions, while Standard
Model states are localized to a 3-brane.  Utilizing compact hyperbolic manifolds
(CHM's) we show that the spectrum of Kaluza-Klein (KK) modes is radically altered.
This allows an early universe cosmology with normal evolution up to
substantial temperatures , and completely negates the constraints on $M_*$ arising from astrophysics.
Furthermore, an exponential hierarchy between the usual Planck scale and the true
fundamental scale of physics can emerge with only ${\cal O}(1)$ coefficients.
The linear size of the internal space remains small.  
The proposal has striking testable signatures.
\end{abstract}
\pacs{PACS:12.10.-g, 11.10.Kk,
\hfill  hep-ph/0002001, CERN-TH-2000-038,\\
11.25.M,04.50.+h
\hfill CWRU-P1-00, SU-ITP-00/05}
\vskip2pc]

Recent work \cite{Anton,ADD,RS,otherearly} has heralded a renewed interest
in higher-dimensional space-times, a  key new concept being the localization
of matter, and even gravity, to branes embedded in the extra
dimensions \cite{polchinski}.  In the canonical example of \cite{ADD},
space-time is a direct product of ordinary 4D space-time
and a (flat) spatial $d$-torus of common linear size $R$
and volume $V_{\rm new} = R^d$, while Standard Model particles
are localized on a 3-brane of thickness $\sim M_*^{-1}$,
where $M_*$ is the new {\em fundamental} higher-dimensional gravitational
(or string) scale.
The low energy effective 4D Planck scale $M_P$ is then
given by the Gauss's Law relation, $M_P^2 = M_*^{2+d}R^d$.
The hierarchy between $M_P$ and $M_*$
can be very large if $R M_*\gg 1$.
For example, if $ d=2$ and $R \sim {\rm mm}$, then $M_* \sim {\rm TeV}$.
The hierarchy $M_P/{\rm TeV}$ thus becomes a problem of understanding
the size of the extra dimensions in such a model \cite{ADMR}.

Remarkably, models with $R$ approaching the sub-millimeter
range are not excluded \cite{ADD3}, but
astrophysics and cosmology do place significant bounds.
In particular, the evolution of the early universe at temperatures just above
those at the epoch of Big Bang Nucleosynthesis (BBN)
is inevitably, and dramatically altered.  This narrow range of normal
evolution prior to BBN makes it difficult to implement baryogenesis, moduli
dilution etc.

The most important model-independent constraints on
such models arise from the production
of light KK modes of the graviton.  These KK modes are the
eigenmodes of the appropriate Laplace operator $\Delta$ on the internal
space, and it is of central importance in the following that
all the constraints
depend on the form of the spectral density
of this operator, which in turn depends completely
on the topology and geometry of the internal space.

In this letter we argue that attractive alternate
choices of compactification imply significantly weaker constraints,
admitting in particular a standard 4D Friedmann-Robertson-Walker (FRW)
evolution
up to high temperatures.  These compactifications
employ a topologically non-trivial internal space---
a $d$-dimensional compact hyperbolic manifold (CHM).
They also throw into a new light the problem
of explaining the large hierarchy $M_P/{\rm TeV}$, since
even though the volume of these manifolds is large,
their linear size $L$ is only slightly larger than the new fundamental
length scale ($L\sim 30 M_*^{-1}$ for example), thus
only requiring numbers of ${\cal O}(10)$.

CHM's are obtained from $ H^{d}$, the
universal covering space of hyperbolic manifolds (those admitting constant
negative curvature), by modding out by an appropriate freely acting discrete
subgroup $ \Gamma  $ of the isometry group of $H^{d}$ \cite{Thurston}.
(If $\Gamma$ is not freely-acting,
then the resulting quotient is a non-flat non-smooth
orbifold. We will not discuss this
interesting case here.)  These manifolds have been much discussed recently as the possible
structure of ordinary 3-space \cite{Starkman}, and play an important
role in the theory of classical and quantum ``chaotic'' systems,
where the spectra of Laplacian operators are also vital \cite{chaos}.
Here we will consider space-times of the form
$M^4\times (H^d/\Gamma|_{\rm free})$ 
($M^4$ is a FRW 4-manifold)
with metric
\be
G_{IJ}dz^I d z^J = g^{(4)}_{\mu\nu}(x)dx^{\mu}dx^\nu + R_c^2 g^{(d)}_{ij}(y)
dy^{i}dy^j.
\label{metric}
\ee
Here $R_c$ is the physical curvature radius of the CHM, 
so that $g_{ij}(y)$ is the metric on the CHM normalized so that
its Ricci scalar is
${\cal R}=-1$, and $\mu = 0,\ldots,3$, $i= 1,\ldots,d$.  

Because they are locally negatively curved, CHM's exist only for $d\ge 2$.
Their properties are well understood only for
$d\leq3$; however, it is known that CHM's
in dimensions $d\geq 3$ possess the important property of
{\em rigidity} \cite{MostowPrasad}.
As a result, these manifolds have {\em no massless shape moduli}.  
Moreover, the volume of the manifold, in units of the curvature radius $R_c$,
cannot be changed while maintaining the homogeneity of the geometry.
Hence, {\em the stabilization of such internal
spaces reduces to the problem of stabilizing a single modulus,}
the curvature length or the ``radion''.
Of course, in a complete high-energy theory, (e.g. string theory),
there will be massive ${\cal O}(M_*)$ excitations of the would-be shape
moduli, and more important for the constraints, the massive KK modes.

To uncover the physics of these models one must consider the spectrum
of small fluctuations $h$ in the metric around the background eq.~(\ref{metric}),
$G_{IJ} \to G_{IJ} + {\rm e}^{ip.x} h_{IJ}(y)$.  There are 3 different types
of KK fluctuations that so arise: $h_{\mu\nu}$, the spin-2 piece;
$h_{ij}$, with indices only in the internal directions, giving
spin-0 fields for the 4D observer; and the mixed case $h_{i\mu}$,
giving spin-1 4D fields.  The 4D KK masses of these states are the eigenvalues
of the appropriate internal-space Laplacians acting on $h(y)$, the correct
Laplacian differing between these 3 cases.
In the most important spin-2
case the operator is the Laplace-Beltrami operator $\Delta_{LB}$
(the Laplacian on scalar functions in the internal space),
defined by
\be
\Delta_{LB} \phi(y) = |g(y)|^{-1/2} \partial_i \left(
|g(y)|^{1/2} g^{ij}  \partial_j\phi(y)  \right) .
\label{beltrami}
\ee

There are no known analytic expressions for the individual
eigenvalues of $\Delta_{LB}$ on a CHM of any dimension.  However,
despite the extremely complicated topology and geometry of CHM's with
arbitrarily large volume, a number of simple facts are generally true.  
First, by a variational argument, the spectrum of
$\Delta_{LB}$ is bounded from below,
and the lowest eigenmode is just the constant function on the CHM.
This zero mode is the internal space wave-function
of the massless spin-2 4D graviton.  As it is a constant,
the effective 4D Planck mass depends only on the {\em volume} of the
(highly curved) internal space.

For example, suppose that the internal space was a 3-sphere
of radius $r$, cut out of an $H^3$ of curvature radius $R_c$.  Its
volume ${\rm Vol}(r)$ grows exponentially for $r\gg R_c$,
\begin{equation}
\label{threevolume}
{\rm Vol}(r)= \pi R_{c}^{3}[\sinh (2r/R_c)-2r/R_c]\, .
\end{equation}
In general, the total volume of a smooth compact hyperbolic
space in any number of dimensions is
\be
{\rm Vol}_{\rm new}=R_c^d~ e^{\alpha }\, ,
\ee
where $\alpha$ is a constant, {\em determined by topology}.
(For $d=3$ it is known that there is a countable infinity of
orientable CHM's, with dimensionless volumes, $e^\alpha$,
bounded from below, but unbounded from above. 
Moreover, the $e^\alpha$ do not become sparsely distributed with large volume.)
In addition, because the topological invariant $e^{\alpha}$ characterizes
the volume of the CHM,
it is also a measure of the largest distance $L$ around the manifold.
CHM's are globally anisotropic; however,
since the largest linear dimension gives the most significant contribution to
the volume, one can employ eq.~(\ref{threevolume}), or its generalizations
to $d\neq3$, to find an approximate relationship between
$L$ and ${\rm Vol}_{\rm new}$.
For $L \gg R_c/2$ the appropriate asymptotic relation,
dropping irrelevant angular factors, is
\be
e^\alpha \simeq \exp\bigl( (d-1) L/R_c \bigr) \ .
\ee

Thus, in strong contrast to the flat case, the expression for
$M_P$ depends {\em exponentially on the linear size},
\be
M_P^2=M_*^{2+d}R_c^d e^{\alpha } = M_*^{2+d} R^d_c \exp\bigl((d-1) L/R_c\bigr) .
\ee
The most interesting case (and as we will see later, most
reasonable) is the smallest possible curvature radius,
$R_c \sim M^{-1}_*$.  Taking $ M_*\sim $~TeV then yields
\be
\label{lmaxnum}
L\simeq 35 M_*^{-1} = 10^{-15} {\rm mm} \, .
\ee
Therefore, one of the most attractive features of a CHM internal space 
is that to generate an exponential
hierarchy between $M_*\sim$~TeV, and $M_P$ requires only that the linear size
$L$ be very mildly tuned.

We now return to the important topic of the non-zero eigenmodes of
$\Delta_{LB}$ on CHM's, and to the astrophysical and cosmological
implications of these KK modes.
Recall that in flat models, the KK modes are extremely light,
$m_{KK}\ge R^{-1} \ge 10^{-4} {\rm eV}$, and very numerous,
$N_{KK} \simeq M_{P}^2/M^2_* \le 10^{32}$ \cite{ADD}.  As a result,
even though these modes are individually only weakly coupled,
with strength $1/M_P$, they can be copiously produced by energetic
processes on our brane, and observational limits then
constrain the fundamental scale.
The tightest astrophysical constraint comes from supernova physics,
leading to a lower bound of $M_* \ge 50 {\rm TeV}$ if $d=2$, and of
$M_* \ge 3 {\rm TeV}$ for $d=3$ \cite{ADD3,KKconstraints}.  There are also
severe limits on the maximum temperature (the ``normalcy temperature''
$T_*$) above which the evolution of the universe must be
non-standard \cite{ADD3}.
This temperature is found by equating the rates for cooling
by the usual process of adiabatic expansion, and by the new process
of evaporation of KK gravitons into the bulk.  This gives
$T_* \le 10$~MeV for $d=2$, up to $T_*\le 10$~GeV when $d=6$.
As we will now see, for us the situation is much improved.

First, by the compactness of the internal space, the
spectrum of $\Delta_{LB}$ on a CHM is discrete and has a gap between
the zero mode and the first excited KK state.
The size of this gap is all important.
Second, most of the eigenmodes of $\Delta_{LB}$ on a
CHM have wavelengths less than
$R_c$, and the number density of these modes is well approximated
by the usual Weyl asymptotic formula
\be
n(k)= (2\pi)^{-d} \Omega_{(d-1)} V_d k^{d-1} \ ,
\label{weyl}
\ee
where $\Omega_{(d-1)}={\rm Area}(S^{d-1})$.
There can also be a few lighter {\em supercurvature modes},
with wavelengths as large as the longest
linear distance in the manifold, and masses thus
bounded below by $L^{-1}$.
There is no simple expression for the spectral density of
supercurvature modes, although the Selberg trace formula
provides some information on the full spectrum of $\Delta_{LB}$.
Nevertheless bounds on the first non-zero eigenvalue
are known.  In the best-studied CHM case of $d=2$
we have the following theorem
\cite{lowest}:
Consider a compact (oriented) Riemann surface $S_g$ of arbitrary genus $g\ge2$,
with metric of constant negative curvature -1.
Then for every $\epsilon$, there exists $N\in Z^+$ such that for
$g>N$ there exists an $S_g$ with first eigenvalue
\be
\lambda_1(S_g) \ge  ( C - \epsilon ) \ ,
\ee
where $C\ge 171/784$ by earlier work \cite{lowest}.  
Restoring units, a large enough volume (and thus genus) $d=2$ CHM
will have first eigenvalue $\ge 171/(784R_c^2)$.  Moreover,
Brooks has conjectured that for $d=2$ a typical
CHM chosen at random will have first eigenvalue $\ge 1/4R_c^2$
with positive probability $P$, perhaps even with $P\to1$
as the genus $g\to\infty$~\cite{Brooks}.  The analogous conjecture 
in $d=3$ is more problematic, but has also been made~\cite{Brooks}.
Numerical studies of the spectra of even small volume $d=3$ CHM's
show that they have very few modes with $\lambda < R_c$ \cite{numeric}.

The crucial result is that the first KK modes are
exponentially more massive than the very light $m_{KK}\ge 1/V^{1/d}$
found in the flat case.  
This drastically raises the threshold for their production.  
Even making the pessimistic assumption that the spectral density
of the supercurvature
modes satisfies eq.~(\ref{weyl}) for $k>1/L$, {\em the astrophysical
bounds of \cite{ADD3} 
and \cite{KKconstraints} completely disappear} since the lightest KK mode
has a mass (at least 30~GeV),
much greater than the temperature of even the hottest astrophysical object.
Similarly the large KK masses imply a much higher normalcy temperature $T_*$,
up to which the evolution of our brane-localized 4D universe can
be normal radiation-dominated FRW.
Approximate numerical evaluation shows that $T_*$
is understandably sensitive to the
gap to the first non-zero KK mass, ranging from 2 GeV to
10 GeV (for $d=2$ to $d=6$)
if $m_{KK,1} \simeq 1/L \simeq~{\rm TeV}/35$, and 
from 20 GeV to 40 GeV if $m_{KK,1} \simeq {\rm TeV}/2$
as suggested by the Brooks conjecture. 
(In all cases taking $M_*=1$~TeV. Raising $M_*$ raises $T_*$.)

So far we have concentrated on the spectrum of $\Delta_{LB}$ appropriate
for the spin-2 KK excitations.  What about the spin-0(1)
excitations?  In both cases the detailed form of the Laplacian
is modified. For example, in the spin-0 case the correct operator is the
Liechnerowicz Laplacian,
\be
(\Delta_{\rm LL} h)_{ij} = -( D^k D_k h_{ij} + R_{ikjl} h^{kl}) ,
\label{lichnerowicz}
\ee
where $D_i$ is the covariant derivative.  
The Mostow-Prasad rigidity theorem for CHM's of dimension
$d\ge 3$ tells us that $\Delta_{\rm LL}$ has no zero modes.  
Although we know of no rigorous bounds for the first eigenvalue
of this operator,
an inspection of the generalized Selberg trace formulae supports
the conjecture that the gap is of similar size to
the Laplace-Beltrami case, a result
that is physically reasonable.  Finally for the spin-1 fluctuations $h_{i\mu}$
recall that these zero modes would correspond to KK gauge-bosons (the original
motivation of Kaluza and Klein!), and are directly related to the continuous
isometries of the compact space.
But, as a result of the quotient by $\Gamma$, CHM's have
no such isometries, and thus there are no massless KK gauge
bosons.  The non-zero
KK modes once again have a mass gap that is at least as large as $1/L$ and is more
likely close to $\sim 1/R_c$, as in the previous cases.
Thus these additional types of fluctuation do not disturb our estimates.

We have not yet addressed
why it is almost automatic that there
exist solutions of the form of eq.~(\ref{metric}).
Since CHM's are just quotients of $H^d$ by a {\em discrete
identification} under $\Gamma\subset{\rm Isom}(H^d)$, it is possible
to find solutions of our form whenever there exists a uniform
negative bulk cosmological constant (CC), given one constraint:
$R_c \sim M_*^{-1}$ and
$e^\alpha \simeq \exp\left((d-1) L/R_c\right)\gg 1$ must
be realized consistently with our ansatz of a factorizable geometry
with a static internal space, {\em together with the vanishing of the
4D long-distance} ($\gg L$) CC.  
To ensure a static internal space,
 the small curvature radius of the internal space must be balanced in the
field equations by the bulk CC, $\Lambda_{4+d} \sim M_*^{4+d}$.
Both these quantities contribute to the
effective long-distance 4D CC, $\Lambda_4$, on our brane,
and typically do not cancel.  Furthermore, one cannot just set
$\Lambda_4$ to zero by adjusting the tension or
energy density $f^4$ of our 3-brane, because this requires
$f^4\gg M_*^4$, violating our basic assumption that a low-energy effective
theory is valid on the brane (and perturbing the geometry, possibly
destroying our assumption that it is factorizable).
To address this problem we must examine the form of the total
4D potential energy density $V$, which in the effective theory
depends only on $R_c$ ($e^\alpha$ is an invariant), and which
arises from the dimensional reduction of the full bulk and brane
actions \cite{ADMR}.

For a 3-brane embedded in $(4+d)$ dimensions,
the bulk and brane actions are respectively:
\begin{eqnarray}
& S_{\rm bulk}  &= \int d^{4+d}x
\sqrt{-|g_{(4+d)}|} \biggl( M_*^{d+2}{\cal R} +\Lambda
- {\cal L}_m \biggr) \\
& S_{\rm brane}  &=  \int d^{4}x \sqrt{-|g_{(4)}^{\rm induced}|}
\biggl( f^{4} + \ldots \biggr),
\label{action}
\end{eqnarray}
where ${\cal L}_m$ is the bulk matter field Lagrangian.
Reduction of these actions gives a
4D potential energy density of the form
\be
V(R_c) = \Lambda R_c^d e^\alpha - M_*^4 e^\alpha (M_* R_c)^{d-2}
+W(R_c) \ ,
\label{potential}
\ee
to which we must add the brane tension $f^4$.
The first two terms arise from the $(4+d)$ bulk CC term,
and the curvature of the internal space.  Now consider
expanding $W(R_c)$, which comes from ${\cal L}_m$,
as a Laurent series in $R_c$
\be
W(R_c)=\sum_{p} a_p \frac{M_*^4}{(R_c M_*)^p} \ ,
\ee
with dimensionless coefficients $a_p$.  (Gauge or scalar
field kinetic energies can give such terms with
$p>0$ \cite{ADMR}.)
If the determination of the minimum is dominated by a competition
between any {\em two} terms in $V$, then at this minimum
$V\equiv V_{\rm min}\neq 0$.  Moreover, $V_{\rm min}$ is enhanced by
$e^\alpha$ over the ``natural'' value $M_*^4$.
However, the vanishing of the 4D CC demands $V_{\rm min}|_{\rm tot}=0$.
This cannot be achieved by adjusting the brane tension
such that $|f^4| \le M_*^4$.

Fortunately there is an attractive alternative.  If {\em three} or more
$R_c$-dependent terms in $V(R_c)$ are all important at the minimum
(for example the CC and curvature terms, and one of the matter
terms from $W$) then we can tune the coefficients $a_p$ such that
$V_{\rm min}=0$, without needing $f^4\gg M_*^4$.  Thus, our basic
assumptions remain consistent.  Moreover, this tuning is
particularly natural in our case precisely because we want the
minimum to occur for a curvature radius close to the fundamental
scale $R_c \sim M_*^{-1}$, at which we expect the high-scale theory to
produce many different terms that contribute roughly in an equal way.
(This is exactly the opposite situation from the large flat extra dimension
case where the minimum has to occur at a length scale much greater than
$M_*^{-1}$.) This one fine-tuning is just the usual
4d CC problem, about which we have nothing to add.

Having shown that there do exist solutions of our form,
another significant result follows from this analysis.
The most severe problem bedeviling the usual large extra
dimension scenario is the radion moduli problem in the early universe
\cite{ADKMR}.  In our case this problem is much weakened.
The radion, which is the light mode
corresponding to dilations of the internal space, gets its mass from
the stabilizing potential $V(R_c)$.  Generally,
in the flat extra dimension scenario, the radion mass $m_r$ is of size
$M_*^2/M_P \simeq 10^{-3}$~eV, so that it is very easily excited
during the exit from inflation.  Furthermore, since its couplings
are $1/M_P$ suppressed, its life-time is longer than the age of the
universe, so that it would unacceptably dominate our current expansion.
In our case, however, the radion mass  is greatly increased because the
second derivative of the potential at its minimum is enhanced by a factor
of $e^\alpha$, $V_{\rm min}''= {\cal O}(e^\alpha M_*^6)$.  Thus
\be
m_r^2 = \frac{1}{2} \frac{R_c^2 V''(R_c)}{e^\alpha M_*^{d+2} R_c^d}
\simeq \frac{1}{R_c^2} \ ,
\ee
which is close to $M_*^2\sim{\rm TeV}^2$.
Therefore, the radion lifetime is
$T \sim M_P^2/M_*^3$, 
much shorter than in the case of flat extra dimensions, 
and only slightly longer than the age of the universe at nucleosynthesis,
even if $M_*\sim {\rm TeV}$.
Moreover, it is (comparatively) easy to dilute away
any unwanted radion oscillations by a period of late inflation.
%n changed the above since the numbers may raise a bit of concern

While cosmologically and astrophysically much safer,
models with internal compact hyperbolic spaces
do have testable signatures accessible to
collider experiments.  Since KK modes 
abound close to the fundamental scale, Standard Model
particle collisions with center-of-mass energies near this scale
will result in the production of KK particles,
detectable by a distinctive missing  energy signature \cite{collider}.
Although this is qualitatively similar to the scenario of \cite{RS},
the spectrum of KK modes one will see is quite distinctive.
While the scale of KK masses is set by $R_c^{-1}$,
their ratios and multiplicities are in almost one-to-one
correspondence with the topology
of the internal manifold \cite{Kac}.
A full exploration of these experimental signatures
will require a more complete investigation of the spectrum of large CHM's,
in particular the issues of isospectrality
and homophonicity of such manifolds.
It is quite likely that such CHM's have other
implications for higher-dimensional physics.
Besides a more detailed study of the question
of radion stabilization, effects such as wavefunction scarring
\cite{chaos} and brane-manifold dynamics are currently under investigation.

We thank L.~Alvarez-Gaume, R.~Brooks, N.~Cornish, S.~Dimopoulos,
C.~Gordon, A.~Gamburd, H.~Mathur, J.~Ratcliffe and  J.~Weeks for discussions, 
and the Stanford (JMR, GDS) and LBNL (JMR) theory groups for hospitality.
Support was provided by the A.P.~Sloan Foundation  (JMR),
the NSF (NK: NSF-PHY-9870115, GDS: NSF-CAREER), and the DOE (GDS, MT).

\end{document}